\documentstyle[aps,prd,preprint]{revtex}

\newcommand{\beq}{\begin{equation}}
\newcommand{\eeq}{\end{equation}}
\def\p{\partial}

\def\lap{\lower.5ex\hbox{$\; \buildrel < \over \sim \;$}}
\def\gap{\lower.5ex\hbox{$\; \buildrel > \over \sim \;$}}
\def\L{\Lambda}
\def\rL{\rho_\Lambda}
\def\rb{\rho_{bare}}
\def\e{\epsilon}
\def\p{\partial}

\begin{document}

\title{Field theory models for variable cosmological constant}

\author{Gia Dvali$^1$ and Alexander Vilenkin$^2$}

\address{
$^1$ Department of Physics, New York University, New York, NY 10003,\\
$^2$ Institute of Cosmology, Department of Physics and Astronomy,
Tufts University, Medford, MA 02155, USA
}

\maketitle

\begin{abstract}

Anthropic solutions to the cosmological constant problem require
seemingly unnatural scalar field potentials with a very small slope or
domain walls (branes) with a very small coupling to a four-form
field.  Here we introduce a class of models in which the smallness of
the corresponding parameters can be attributed to a spontaneously
broken discrete symmetry.  We also demonstrate the equivalence of
scalar field and four-form models.  Finally, we show how our models
can be naturally embedded into a left-right extension of the standard
model.

\end{abstract}

\section{Introduction}

Particle physics models suggest that natural values for the
cosmological constant $\L$ are set by a high-energy scale, anywhere
between 1 TeV and the Planck scale $M_P$ (for reviews see
\cite{Weinberg,Sahni,Carroll}).  The corresponding vacuum energy
density $\rL$ is then between $(1 ~TeV)^4$ and $M_P^4$.  On the
other hand, recent observations indicate \cite{data} that the actual
value is $\rL\sim (10^{-3}~eV)^4$, at least 60 orders of magnitude
smaller.  It is hard to explain why $\rL$ should be so small.  Even
more puzzling is the fact that $\rL$ appears to be comparable to the
present matter density $\rho_{m0}$.  The two densities scale very
differently with the expansion of the universe, and it is very
surprising that they nearly coincide at the present time.

To our knowlege, the only approach that can explain both of these
puzzles is the one that attributes them to anthropic selection effects
\cite{W,AV95,Efstathiou,MSW,GLV,GV,Bludman,Weinbergcomment,Donoghue,GV2}.
In this approach, what we perceive as the cosmological
constant is in fact a variable that can take different values in
different parts of the universe and it is assumed that the fundamental
theory allows such a variation of the effective $\L$.
However, the particle physics models of variable $\L$ suggested so far
appear to have some unnatural features.

In one class of models, the role of $\rL$ is played by a slowly
varying potential $V(\phi)$ of some scalar field $\phi$ which is very
weakly coupled to ordinary matter.  The simplest example is
\beq
V(\phi)=\rb +{1\over{2}}\mu^2\phi^2,
\label{quadratic}
\eeq
where $\rb$ is the ``true'' cosmological constant and it is assumed
that $\rb$ and $\mu^2$ have opposite signs.  The two terms on the
right-hand side of (\ref{quadratic}) nearly cancel one another in
habitable parts of the universe.  In order for the evolution of $\phi$
to be slow on the cosmological timescale, the mass parameter $\mu$
has to satisfy the condition \cite{GV}
\beq
|\mu|\ll 10^{-120}M_P^3|\rb|^{-1/2}.
\label{mubound}
\eeq
The challenge here is to explain this exceedingly small mass scale in
a natural way.  The possibilities suggested so far include a
pseudo-Goldstone field $\phi$ which acquires a potential through
instanton effects \cite{GV,GV2}, a large running of the field
renormalization \cite{Weinbergcomment}, and a non-minimal kinetic term with
an exponential $\phi$-dependence \cite{Donoghue,GV2}.

In the above mentioned class of models, the vacuum energy density
$\rL$ takes values in a continuous range.  An alternative possibility
is that the spectrum of $\rL$ is discrete, as in the ``washboard''
potential model suggested in the early paper by
Abbott \cite{Abbott}.  A simple example of such a potential is
\beq
V(\phi)=-A\cos(2\pi\phi/\eta)+\epsilon\phi/\eta.
\label{cos}
\eeq
For $\epsilon\ll 2\pi A$, the
potential has local minima at $\phi_n\approx n\eta$ with $n=0,\pm 1,\pm2,
...$, separated from one another by barriers.  The vacuum at
$\phi=\phi_n$ has energy density
\beq
\rho_{\L n}\approx n\epsilon.
\label{rhon}
\eeq
Transitions between different vacua can occur through bubble
nucleation.

Another version of the discrete model, first discussed by Brown and
Teitelboim \cite{BT}, assumes that the cosmological constant is due to
a four-form field \cite{Duff},
\beq
F^{\alpha\beta\gamma\delta}={F\over{\sqrt{-g}}}\e^{\alpha\beta\gamma\delta},
\label{F}
\eeq
which can change its value through nucleation of branes.  The total
vacuum energy density is given by
\beq
\rL=\rb +F^2/2
\label{rhoF}
\eeq
and it is assumed that $\rb<0$.  The change of the field across the
brane is
\beq
\Delta F=q,
\label{Fq}
\eeq
where the ``charge'' $q$ is a constant fixed by the model.  The four-form
model has recently attracted much attention
\cite{Bousso,Donoghue,FMSW,Banks,GV2} because four-form fields
coupled to branes naturally arise in the context of
string theory.

In the range where the bare cosmological constant is almost
neutralized, $|F|\approx |2\rb|^{1/2}$, the spectrum of $\rL$ is
nearly equidistant, with a separation
\beq
\Delta\rL\approx |2\rb|^{1/2}q.
\label{Deltarho}
\eeq
In order for the anthropic explanation to work, $\Delta\rL$ should not exceed
the present matter density,
\beq
\Delta\rL\lesssim\rho_{m0}\sim (10^{-3}~eV)^4.
\label{ebound}
\eeq
With $\rb\gtrsim (1~TeV)^4$, it follows that
\beq
q\lesssim 10^{-90}M_P^2.
\label{qbound}
\eeq
Once again, the challenge is to find a natural explanation for such
very small values of $q$.

Feng, March-Russell, Sethi and Wilczek (FMSW) \cite{FMSW} have argued that the
required small values of the charge $q$ can naturally arise due to
non-perturbative effects in M-theory. Their assumption is
that some of the fundamental string theory branes can have an
extraordinarily small tension $\sigma \lesssim 10^{-90}M_P^3$.
The idea was to use D2-branes obtained by wrapping of $k$ world-volume
coordinates of a higher-dimensional
Dp-branes ($p= 2 + k$) on a collapsing $k$-cycle. The volume of the cycle
$V_k$ then determines the effective tension of the resulting 2D-brane in
lower dimension $\sigma \sim V_k$.
It was assumed that non-perturbative
quantum corrections may stabilize the volume
at an exponentially small size, resulting
in an exponentially small
2-brane tension.
Then, assuming
a typical tension-to-charge relation,
\beq
q \sim {\sigma\over{M_P}},
\label{tq}
\eeq
one arrives
at the required value (\ref{qbound}) of the brane charge.

The problem with this approach is that the small brane tension is not
protected against quantum corrections below the supersymmetry breaking
scale.  Branes of the kind discussed by Feng {\it et. al.} originate
as solitons of the fundamental theory, valid above the field theory
cutoff $M$.\footnote{In our discussion everywhere $M$ should be understood
as the ultraviolet cut-off of the field theory, for which we take the string
scale. We shall assume that the string coupling is of order one and
none of the string compactification radii are large.
Thus the theory below $M$ is an effective four-dimensional
theory and roughly $M_P \sim M$.}
As a result,
in the low energy effective field theory valid below the scale ``M''
these branes behave as
fundamental objects, in the sense that the low energy observer cannot
``resolve'' their structure.
Thus, the world-volume theory of these branes is a $2+1$-dimensional field
theory with a cut-off $\sim M$.
In the absence of supersymmetry
such branes would be expected to have a
tension $\sigma \sim M^3$  and the assumption of $\sigma \ll M^3$
would be extremely unnatural.
The existence of a spontaneously broken low-energy
supersymmetry may ameliorate the situation a bit, but unfortunately
not up to a sufficient level, as we shall argue below.

The problem is
that, in the absence of an exact supersymmetry, the brane world-volume
states will induce an unacceptably high
brane tension through quantum loops.
To demonstrate this, let us assume the most optimistic scenario,
when the brane sector is
only gravitationally coupled to the sector that spontaneously breaks
supersymmetry. The lowest possible scale of supersymmetry breaking
compatible with observations is somewhere around TeV energies.
Thus, the mass splitting among the brane superfields induced by gravity will
be at least
\beq
m^2_S \sim TeV^4/M_P^2
\eeq
(this is the usual magnitude for the gravity-mediated
supersymmetry breaking). Each brane superfield will induce a
linearly-divergent contribution to the brane tension. For instance,
at one-loop the contribution coming from a massless supermultiplet
is
\beq
 \Delta\sigma \sim m_S^2\int {d^3p \over p^2 - m_S^2}
\sim m_S^2 \Lambda_{cut-off}
\eeq
 which we (at best)
can cut off at the scale $M\sim M_P$.
Thus, the resulting contribution to the brane tension
from each pair of modes is expected to be
\begin{equation}
   \Delta\sigma = TeV^4/M_P.
\label{correction}
\end{equation}
Thus, a single world-volume supermultiplet already gives an unacceptably
large renormalization of the brane tension.

In addition, we have to sum over all the world-volume
modes with masses below $M$. In the absence of an explicit model,
it is hard to argue what the precise density of the modes is, and we will not
speculate further on this issue.
However, one may expect that an additional
enhancement factor, can be as large as $\sim M/\sigma^{1/3}$
(since we expect the spacing of the modes in the world-volume theory to be
set by the brane tension, the only scale in the low energy world-sheet
Lagrangian). This would give the resulting tension to be something like
$\sigma\sim TeV^3$.  Although we cannot exclude a
miraculous cancellation among the different modes, such a cancellation
is not indicated by any symmetry and would be hard to understand in an
effective field theory picture. It is unlikely that string theory corrections
can cure the problem since they set in only above the scale $M$, and
such a conspiracy among high and low energy physics would constitute
a violation of the decoupling principle.

From the above arguments, we expect that branes with very low tension
and charge should be looked for among the effective field theory solitons.
The tension of such solitonic branes is much better protected from
quantum destabilization.
As a simple straightforward example let us construct
a solitonic brane with tension $\sigma\sim TeV^6/M^3$.
Let $\phi$ be a chiral superfield with a superpotential
\begin{equation}
  W = {\phi^3 \over 3}
\end{equation}
As above, let us assume that supersymmetry is spontaneously broken
at the TeV scale in a sector that couples to $\phi$ only gravitationally.

To be more explicit, let $S$ be a superfield which spontaneously breaks
supersymmetry through the expectation value of its auxiliary
($F_S$)-component.
By assumption $F_S \sim TeV^2$, which is the lowest possible
phenomenologically acceptable scale. Then according to the standard picture,
SUSY breaking in the $\phi$-sector will be transmitted through
gravity via couplings of the form
\beq
\int d^4\theta {S^*S \over M_P^2}\phi^*\phi +
\int d^2\theta {S \over M_P}\phi^3  +..
\eeq
where $\theta$ is a superspace variable.
The resulting soft mass of the $\phi$ scalar is $m^2_S \sim
{|F_S|^2 \over M_P^2} \sim TeV^4/M_P^2$
and the sign
depends on the details of the theory. We shall assume that the sign
is negative. Then the effective potential for $\phi$ can be written as
\begin{equation}
V(\phi) = -m^2_S \phi^*\phi - (cm_S\phi^3 + {\rm h.c.}) + |\phi|^4,
\end{equation}
where $c$ is a number of order one. Obviously, this system exhibits
a spontaneous breaking of the discrete $Z_3$ symmetry
($\phi \rightarrow {\rm e}^{i{2n\pi \over 3}}\phi$)
and admits topologically stable solitonic
brane solutions (domain walls) with tension $\sigma \sim m_s^3 \sim TeV^6/M^3$.
These branes do not suffer from the quantum correction problems
discussed above, due to the fact that all the integrals in the brane
world volume theory
are cut off at the compositness scale, which coincides with the brane
tension scale $m_s$. Above this scale, one has to do computation in the
full theory in which there is no renormalization of the mass parameters
beyond the scale $m_S$, due to low-scale supersymmetry.

 Although the above simple example demonstrates that the field theory branes
with low tension are easily possible, it does not achieve our primary goal of
generating very low charge branes, since the above branes are not coupled
to any four-form field.\footnote{It was shown in
\cite{susy} that domain walls of
large-$N$ supersymmetric $SU(N)$-gluodynamics are automatically charged
with respect to a (composite)
three-form field. These are $Z_N$-walls formed by the gluino condensate,
and both the tension and the charge of these objects
are set by the $SU(N)$-QCD scale $\Lambda$.
However, there are two points that probably make these walls useless
in the present context.  First, the four-form field strength is the same
on both sides of the wall, due to the fact that its change gets compensated
by the change of the phase of gluino condensate.
Second, these walls usually cannot
exist if the supersymmetry breaking scale is $> \Lambda$.
Therefore, according to our previous arguments, their charge can at best be
$\sim TeV^4/M_P^2$, which is too high for our purposes.}
This will be our task in the following
discussion.

Thus, in this paper we report on our search for field-theoretic models in
which the smallness of $q$ (or of $\mu$ in Eq. (\ref{quadratic})) can be
attributed to some symmetry.
We found a class of models in which the value of $q$ (or
$\mu$) can be made arbitrarily small.  The charge-to-tension ratio
$q/\sigma$ can also be made as small as desired.
As a byproduct of this research, we
found a somewhat unexpected equivalence relation between scalar field and
four-form models.

After outlining the general idea in Section II, we shall first
discuss, in Section III, a simplified
$(1+1)$-dimensional version of our model.  The equivalence between
four-form and scalar field models is demonstrated in Section IV.
Models in $(3+1)$ dimensions are discussed in Section V.  In Section
VI we illustrate how our models can be naturally embedded into a
left-right symmetric extension of the standard model. Our
conclusions are briefly summarized and discussed in Section VII.

\section{general philosophy}.

 Our idea is the following. 1) Branes with an extremely low four-form charge
can appear in the form of solitons (domain walls) of an effective low energy
theory.
2) The small value of the charge-to-tension ratio is natural in the sense
that it can be arbitrarily suppressed by the symmetries of the model.

 Before proceeding to specific examples, let us briefly discuss the
main ingredients of our scheme: 1) a real scalar field,
$a$, which can be thought of as the phase of a certain
complex scalar field $X$ with a nonzero vacuum expectation value
$\langle X \rangle$;
2) a four form field $F_{\mu\nu\sigma\tau}$ which can be obtained
from a three-form potential, $F_{\mu\nu\sigma\tau}=\p_{[\mu}
A_{\nu\sigma\tau]}$;
3) a scalar field $\phi$, which spontaneously breaks a $Z_{2N}$
discrete symmetry at the scale $\langle\phi\rangle$.
The crucial assumption is that both scales $\langle\phi\rangle$
are $\langle X \rangle = \eta$ are well below the cutoff scale  $M$.
Having in mind a low energy SUSY, their natural value can be as low as
TeV, which we shall adopt for definiteness\footnote{
In practice, even much higher scales can do the job, provided $N$ is chosen to
be large enough.}.

We require that the action be invariant
under the following three symmetries:
1) $Z_{2N}$ symmetry under which
\beq
\phi \rightarrow \phi e^{i\pi/N}, ~~~~~a \rightarrow -a,
\label{Z2N}
\eeq
2) symmetry under the shift
\beq
a \rightarrow a+2\pi\eta,
\label{shift}
\eeq
and 3) the three-form gauge transformation
\beq
 A_{\mu\nu\alpha} \rightarrow A_{\mu\nu\alpha} +
\partial_{[\mu}B_{\nu\alpha]}
\eeq
where $B_{\nu\alpha}$ is a two-form.
In addition, we shall assume that there is an (at least) approximate
global $U(1)$ symmetry $X \rightarrow e^{i\theta}X$, so that it is meaningful
to talk of $a$ as the phase degree of freedom. Thus, $a$ can be regarded as
a sort of a (pseudo)Goldstone particle.
Note that even if $U(1)$ is explicitly broken by some non-perturbative
Planck-scale-suppressed corrections,
the mass of $a$ will be suppressed by the powers of
$\eta/M \sim TeV/M$,
and is much smaller than the masses of other scalars in the theory,
which we assume are around the TeV scale.
Thus, below TeV energies we can
integrate out the heavy quanta, such as $\phi$ and the radial part of $X$,
and derive an effective low energy action for the remaining light
fields, $A_{\mu\nu\alpha}$ and $a$. We shall assume that the
low energy action includes all possible interactions
that are compatible with the unbroken symmetries. Obviously,
operators that are forbidden by
spontaneously broken symmetries must appear suppressed (at least) by powers
of the corresponding Higgs VEVs. Among such operators there is
a mixing of $a$ with the three-form potential,
\begin{equation}
\epsilon^{\mu\nu\alpha\beta}
A_{\mu\nu\alpha}\partial_{\beta}a.
\label{mixingaA'}
\end{equation}
Since at high energies this operator is forbidden by the
$Z_{2N}$-symmetry (as well as by an approximate global $U(1)$-symmetry),
in the low energy theory it should
appear suppressed by the following factor
\begin{equation}
{\langle\phi^N\rangle\over M^N}\epsilon^{\mu\nu\alpha\beta}
A_{\mu\nu\alpha}\partial_{\beta}a + {\rm h.c.}
\label{mixingaA}
\end{equation}
Power $N$ is dictated from the fact that $\phi^N$ is the lowest possible
power of $\phi$ that makes up $Z_{2N}$-invariant in combination with $a$.
Thus, the mixing can be arbitrarily suppressed by powers of
$TeV/M$ due to the symmetry reasons. This is enough to realize our program:
the extremely small mixing coefficient
automatically translates into an extremely small four-form charge
of $a$-field domain walls. Like axionic domain walls,
these walls must be present due to the $2\pi$ periodicity of the
$a$-potential.  Since $a$ changes by $2\pi$ across the wall, each wall
acts as a source for the four-form field. The resulting four-form charge
is suppressed by the $a-A$ mixing
(\ref{mixingaA}) and is miniscule. We shall discuss this in more detail below.

Before proceeding, let us make a brief note.
Below the energies comparable to the masses
of either $\phi$ or the radial $X$ quanta, $\langle\phi\rangle$ and
$|\langle X\rangle|$ can be regarded as constants and
the coupling (\ref{mixingaA})
is a local gauge-invariant operator.
For higher energies, however, one has to include
additional momentum-dependent interactions.
These become relevant for processes with
external $\phi$ and $X$ legs and must
be included because of gauge invariance.
Any interaction that will be responsible for inducing the above
mixing term in the low energy theory will also generate
momentum-dependent operators required by this gauge invariance
(see the next section). These additional interactions, however,
will play no role in our
analysis, since they only contribute
to processes with $\phi$ and $X$-quanta emission that are
forbidden at energies of our interest.

\section{A toy model in $(1+1)$ dimensions}

The field content of our $(1+1)$-dimensional toy model is:
the ``electromagnetic'' vector potential $A_{\mu}$,
a real scalar field $a$, an
electrically charged
(Dirac) fermion $\psi$, and a
complex scalar field $\phi$.
The Lagrangian is
\begin{eqnarray}
L &=& i\bar{\psi}\gamma_{\mu}D^{\mu}\psi + \bar{\psi}\gamma_{\mu}\psi
{\phi^N\over M^N}\epsilon^{\mu\nu}\partial_{\nu}a +
{1\over{2}}(\partial_{\mu} a)^2 + {1\over{2}}(\partial_{\mu} \phi)^2
\nonumber\\
&-& V(a,\phi) - {1 \over 4}F_{\mu\nu}F^{\mu\nu}
+ higher~derivative~terms,
\label{L2}
\end{eqnarray}
where $D^{\mu}$ is the usual covariant derivative with respect to
the electromagnetic  $U(1)$ group, $V(a,\phi)$ is
some potential function, and $1/M$ is a constant of order one.
Note that in $(1+1)$ dimensions $M$ must be dimensionless, since the canonical
dimensionalities of $A_{\mu}$, $\phi$ and $a$ are zero.  For
simplicity, here and below we use flat spacetime metric.  Our sign
conventions are $g_{00}=\e^{01}=1$.

 We must stress that the interaction with fermions is introduced
exclusively for the
illustrative purpose. It allows us to trace explicitly how the
gauge-invariant mixing operator arises in perturbation theory.
In reality we will not necessarily rely on fermions,
but assume that the mixing is induced by some perturbative or
non-perturbative
physics at the cut-off scale.

As before, we require that the action be invariant
under the following three symmetries:
1) $U(1)$-gauge invariance, 2) $Z_{2N}$ symmetry (\ref{Z2N}),
and 3) the shift symmetry
(\ref{shift}).
Under these symmetries the function $V(a,\phi)$ is
determined to depend on $a$ and $\phi$ only through the invariants
$\cos(a)$, $\phi^{2N}$ and $\phi^N\sin(a)$.
The precise form of this function will be of no
importance for us, as long as
the vacuum expectation value $\langle\phi\rangle$ of $\phi$ is nonzero and
$\langle\phi\rangle << M$.
This expectation value spontaneously breaks
$Z_{2N}$ symmetry down to nothing, and some operators forbidden by this symmetry
will be generated with strength suppressed by powers of the ratio $\epsilon=
{\langle\phi\rangle \over M}$.

Among such operators we shall be interested in $A_{\mu}$-$a$ mixing,
which appears as a result of one-loop fermionic exchange. The corresponding
operator has the form\footnote{We thank A. Grassi for a clarifying discussion
on this issue.}
\begin{equation}
g{\phi^N\over M^N}\epsilon^{\mu\nu}\partial_{\mu}a(g_{\nu\rho} -
{\partial_{\nu}\partial_{\rho} \over \partial^2})A^{\rho},
\label{operator}
\label{O}
\end{equation}
where $g$ is a loop factor that includes a dimensionful gauge coupling.
Shifting the $\phi$ field, we shall expand the theory around the vacuum
state: $\phi\rightarrow \langle\phi\rangle + \phi(x_{\mu})$.
Performing integration by parts in the first term
of the expansion, Eq. (\ref{O}) can be written as
\begin{equation}
g{\epsilon^N}\epsilon^{\nu\mu}A_{\mu}\partial_{\nu}a
+ g\left({N\epsilon^{N-1}\phi(x_{\mu})\over M^N} + ...\right)
\epsilon^{\mu\nu}\partial_{\mu}a(g_{\nu\rho} -
{\partial_{\nu}\partial_{\rho} \over \partial^2})A^{\rho},
\label{exp}
\end{equation}
where ellipses stand for the terms that contain higher powers of
$\phi(x_{\mu})$. These terms describe $A_{\mu}$-to-$a$
transition via emission of $\phi$-particles, and therefore are not
relevant at the energies below
the mass of the $\phi$-quanta. Thus, at low energies
the only relevant operator is the first term in (\ref{exp}).
An important fact is that this term is parametrically suppressed by
powers of small quantity $\epsilon$.
After integrating out the heavy fields and the fermions, the relevant part
of the low energy Lagrangian can be written as
\begin{equation}
L = {1\over{2}}(\partial_{\mu} a)^2 - {1 \over 4}F_{\mu\nu}F^{\mu\nu}
- V(a) +
g{\epsilon^N}\epsilon^{\nu\mu}A_{\mu}\partial_{\nu}a
+ higher ~derivative~ terms
\label{effect}
\end{equation}

The potential $V(a)$ in (\ref{effect})
is the effective potential for $a$. As indicated above,
the shift symmetry requires it to be an arbitrary periodic function
of $a$. In the absence of $a-A$ mixing, the theory would have an
infinite set of degenerate vacua at the minima of this potential,
and standard topological arguments would
imply that there must be domain wall
configurations with $a$ changing by $\Delta a=2\pi$ across the
wall.

Let us now see how this situation is affected by the $a-A$ mixing term
in (\ref{effect}).  The gauge field strength can be expressed as
\beq
F^{\mu\nu}=\e^{\mu\nu}F.
\label{Fe}
\eeq
The equation of motion for $A$ field
\begin{equation}
 \partial_{\nu}F^{\mu\nu} =
-g{\epsilon^N}\epsilon^{\mu\nu}\partial_{\nu}a
\label{Feq}
\end{equation}
then implies that the change of $a$ across the wall is accompanied
by a change of the field strength,
\beq
\Delta F =-g{\epsilon^N}\Delta a.
\eeq
The vacua on the two sides of the wall will not generally be
degenerate, due to this difference in the field strength.
Higher-energy vacua will decay into lower-energy ones through
nucleation of bubbles with the domain walls at their boundaries.

We thus see that
the walls have acquired a charge
\beq
q=2\pi g\e^N,
\label{qwall}
\eeq
which is suppressed by a power of the small parameter $\e$.
A remarkable property of the above model is that, with a suitable
choice of $N$, this charge can be made
arbitrarily small.

\section{Equivalent scalar field model}

The scalar field equation, obtained by varying Eq. (\ref{effect})
with respect to $a$, is
\beq
\p^2 a+V'(a)-g\e^N F=0.
\label{aeq}
\eeq
Now, the field equation (\ref{Feq}) for $F$ can be integreated to
yield
\beq
F=-g\e^N(a-a_0),
\label{Fsol}
\eeq
where $a_0$ is an integration constant.  Substituting this into
(\ref{aeq}), we see that the resulting equation for $a$ is that for a
scalar field with a potential
\beq
U(a)=V(a)+{1\over{2}}\mu^2(a-a_0)^2,
\label{U}
\eeq
where
\beq
\mu=g\e^N.
\label{mu}
\eeq
Let us recall that $V(a)$ is periodic in $a$, $V(a+2\pi)=V(a)$.
Hence, the potential $U(a)$ is a ``washboard'' potential of the kind
considered by Abbott \cite{Abbott}.

The energy-momentum tensor for our model is
\begin{eqnarray}
T_{\mu\nu} &=& \p_\mu a\p_\nu a-{1\over{2}}g_{\mu\nu}\p_\sigma a\p^\sigma
a +g_{\mu\nu}V(a)+{1\over{2}}g_{\mu\nu}F^2 \nonumber\\
&=& \p_\mu a\p_\nu a-{1\over{2}}g_{\mu\nu}\p_\sigma a\p^\sigma
a+g_{\mu\nu} U(a),
\label{Tmunu}
\end{eqnarray}
where
we have used Eq. (\ref{Fsol}).  The last expression in
(\ref{Tmunu}) coincides with the energy-momentum tensor for a scalar
field with a potential $U(a)$.  We thus see that, as long as
non-gravitational interactions of $a$ and $A_\mu$ are negligible
(apart from their interaction with one another), the model
(\ref{effect}) is equivalent to that of a scalar field with a
potential $U(a)$ given by Eq. (\ref{U}).  Tunneling between vacua with
different values of $F$ is replaced in this scalar model by tunneling
between different minima of the washboard potential.

\section{Models in $(3+1)$ dimensions}

We can easily generalize our model to $3+1$ dimensions.
The main difference is that instead of the vector potential we shall consider
a three-form field $A_{\mu\nu\alpha}$ and require the gauge invariance
under $A_{\mu\nu\alpha} \rightarrow A_{\mu\nu\alpha} +
\partial_{[\mu}B_{\nu\alpha]}$.
The Lagrangian of interest then becomes
\begin{equation}
L = {1\over{2}}\eta^2(\partial_{\mu} a)^2 - {1 \over 4}F^2
- V(a) + g\eta^2{\phi^N\over M^{N}}{\tilde A}^\mu\partial_{\mu}a,
\label{L4}
\end{equation}
where
\beq
{\tilde A}^\mu =\epsilon^{\mu\nu\sigma\tau}A_{\nu\sigma\tau},
\eeq
$a$ is dimensionless and $\eta$ has dimension of energy (it has
the meaning of the shift symmetry breaking scale).
The last term in (\ref{L4}) should be understood as an effective
low-energy interaction arising from some fundamental theory, with
$M$ being understood as the Planck scale.  It is the lowest-dimension
operator consistent with gauge, shift and $Z_{2N}$ symmetries.

The shift symmetry (\ref{shift}) suggests
that $a$ might be an axion-like field arising as a phase of some
complex scalar, $X=Re^{ia}$.  Then $|\langle X\rangle | =\eta$ and
the $Z_{2N}$ symmetry is
\beq
\phi\rightarrow\phi e^{i\pi/N},~~~~~~ X\rightarrow X^{\dagger}.
\label{Z2N4}
\eeq
In terms of the fields $X$ and $\phi$, the last term in (\ref{L4}) can
be written as
\beq
ig(\phi/M)^N{\tilde A}^\mu (X\p_\mu X^{\dagger} -X^{\dagger}\p_\mu X).
\eeq

We shall assume
that the $Z_{2N}$ symmetry is spontaneously broken at a scale
much below $M$. Having in mind a low energy supersymmetry, this breaking
scale may be as small as TeV, without fine tuning.
This would imply that the parameter $\epsilon =\langle\phi\rangle /M
\sim 10^{-15}$.

As before, at low energies we can keep only a constant part in $\phi$,
in which case the last term in (\ref{L4}) reduces to
\begin{equation}
g\eta^2{\epsilon^N}{\tilde A}^\mu
\partial_{\mu}a.
\end{equation}
The field equation for the four-form field (\ref{F}) is
\beq
\p_\mu F=g\eta^2 \e^N\p_\mu a,
\label{Feq4}
\eeq
and we obtain
\beq
\Delta F= g\eta^2 \e^N\Delta a.
\label{DeltaF4}
\eeq
With a periodic potential $V(a)$, the model has domain wall solutions
with $a$ changing by $\approx 2\pi$ across the wall, and
Eq. (\ref{DeltaF4}) indicates that these domain walls acquire a charge
\beq
q=2\pi g\eta^2 \e^N.
\label{q4}
\eeq
For $\e\sim 10^{-15}$ and $\eta\lesssim M$, the condition
(\ref{qbound}) is satisfied with $N\geq 6$.

With $\e$ so small, the effect of the field $F$ on the structure of
domain walls is negligible.  The wall tension $\sigma$ is determined solely by
the potential $V(a)$ and is not suppressed by powers of $\e$.  This is
in contrast to M-theory based models, where $q$ and $\sigma$ are
related by (\ref{tq}).

Following the same steps as in Section III, it can be shown that our
model is equivalent to a scalar field model with a potential
$U(a)$ given by (\ref{U}),
\beq
U(a)=V(a)+{1\over{2}}\mu^2(a-a_0)^2,
\label{U4}
\eeq
where now
\beq
\mu=g\e^N \eta.
\label{mu4}
\eeq
With $\mu$ suppressed by powers of $\e$, this washboard potential is
of the type required in the Abbott's model \cite{Abbott}.

An interesting version of the model is obtained by replacing the
discrete shift symmetry by a symmetry with respect to arbitrary
translations,
\beq
a\rightarrow a+{\rm const}.
\label{shift4}
\eeq
Then $a$ is a Goldstone boson and $V(a)=0$.  The equivalent scalar
model has the potential
\beq
U(a)={1\over{2}}\mu^2(a-a_0)^2.
\eeq
This is of the same form as in the simple model (\ref{quadratic}) in
which the effective vacuum energy density takes values in a continuous
range.  With $|\rb|\sim (1 ~TeV)^4$, the condition (\ref{mubound}) on
$\mu$ is satisfied for $N\geq 6$.

\section{Embedding into Particle Physics Models}.

 In this section we would like to show that our model can be naturally embedded
in well-motivated particle physics models. As an example we shall consider an
embedding into a left-right symmetric extension of the standard model.
We will see that the breaking of $Z_{2N}$ symmetry can be associated with the
spontaneous breaking of a left-right symmetry. We shall start
by briefly reviewing a minimal left-right symmetric extension of the
standard model\cite{lr}
 The gauge group is $SU(3)\otimes SU(2)_L\otimes SU(2)_R \otimes U(1)\otimes P$
where $P$-parity interchanges left and right $SU(2)$-subgroups. This
will be identified with our parity symmetry.
Left handed and right handed fermions form doublets of $SU(2)_L$ and
$SU(2)_R$ groups respectively. Fermion masses are generated via coupling to
a by-doublet Higgs field (doublet under both $SU(2)$-s) which we shall
denote by $H$. From the point of view of the standard
model subgroup, this by-doublet field includes two electroweak
Higgs doublets $H_U$ and $H_D$,
which give masses to ``up'' and ``down'' quarks (and leptons) respectively.
For instance, the Yukawa couplings that give masses to
quarks $Q_{L,R} = (U_{L,R}, D_{L,R})$
\begin{equation}
  H\bar{Q_L} Q_R = H_U\bar{Q_L} U_R  + H_D\bar{Q_L} D_R,
\end{equation}
where we have written the $SU(2)_L\times U(1)_Y$-decomposition.
This coupling has an accidental Peccei-Quinn type global symmetry. The
pseudo-Goldstone boson of this would-be symmetry is the phase difference
of the electrically-neutral
components of $H_U$ and $H_D$ doublets. This pseudoscalar state changes
the sign under $P$-parity transformation.
However, since this global symmetry is not exact,
the would-be Goldstone boson is massive. Below our $a$ field
will be identified with this boson.

The left-right symmetry can be broken by introducing two doublet Higgs fields,
$H_L$ and $H_R$, that are doublets under left and right $SU(2)$s respectively.
One of them must develop an expectation value around the TeV scale (or above)
thereby breaking the corresponding $SU(2)$ and left-right symmetry.
The remaining
$SU(2)$ subgroup then has to be identified with the electroweak
$SU(2)_L$ group of the standard model.

A symmetry breaking in which only one of the doublets gets an
expectation value can be achieved by an appropriate choice of the
parameters in the Higgs potential.  An important point, however,
is that the presence of quartic terms in the potential is
essential for the symmetry breaking.
In a supersymmetric theory, such terms cannot be
written at the renormilizable level, and as a result one has to introduce
additional singlets, that are parity odd.\footnote{Note that without
such singlet
fields, one would have to include non-renormalizable quartic couplings
suppressed by $M_P$ in the superpotential. These, however,
will create asymmetric minima at a
very high scale, which is incompatible with the phenomenologically
most interesting low energy LR-symmetric extension.}

The role of one such signlet
in our case will be played by the $\phi$ field. The relevant couplings in the
potential have the form
\begin{equation}
   {\phi^{N} \over M^{N-2}} (H_L^*H_L - H_R^*H_R)
-m^2(H_L^*H_L + H_R^*H_R) +
\lambda (H_L^*H_L + H_R^*H_R)^2 + \lambda'(H_L^*H_LH_R^*H_R)
\label{RL}
\end{equation}
where $M, m$ are the mass scales around TeV and $\lambda, \lambda'$ are
positive constants. There are many other possible terms that are not essential
for our discussion. The only requirement to the rest of the potential
is that it forces $\phi$ to get a nonzero expectation value. Then, since the
coupling with $\phi$ creates a left-right asymmetry, only one of the
doublets is expected to get a VEV.

The coupling
between $\phi$ and the doublets in (\ref{RL}) is non-renormalizable. However,
it can be easily
obtained by integrating out additional singlet fields of mass
$M$ and zero VEV. For instance, for $N=6$ it is enough to introduce a single
scalar field $\chi$ that under $Z_{12}$ symmetry transforms as
$\chi \rightarrow {\rm e}^{i\pi/2}\chi$
\begin{equation}
  \phi^3 \chi + \phi^{*3}\chi^* +
 M^2\chi^*\chi + \chi^{*2} (H_L^*H_L - H_R^*H_R)
\end{equation}
Integrating out $\chi$ we arrive at the effective non-renormalizable
coupling given in Eq. (\ref{RL}).
Below the TeV scale, the remaining effective
low energy theory is the standard model
coupled to our field $a$, via the pseudoscalar interaction. All other
ingredients of our scheme (e.g. mixing with a three-form, etc.)
are assumed to be unchanged.
This completes the embedding of our model into a $LR$ symmetric extension
of the standard model.

\section{Discussion}

As explained in the Introduction, anthropic solutions to the
cosmological constant problems require scalar field potentials with a
very small slope or domain walls (branes) with a very small coupling
to a four-form field.  Here we introduced some models in which the
smallness of the corresponding parameters can be attributed to a
$Z_{2N}$ symmetry, (\ref{Z2N}) or (\ref{Z2N4}).

We note that, apart from the desired domain walls with small coupling
to the four-form field, our models may have a variety of other
topological defects.  The breaking of the discrete $Z_{2N}$ symmetry
is accompanied by the formation of $\phi$-walls such that the value of
$\phi$ changes by a factor of $e^{i\pi/N}$ across the wall.  $2N$ such
walls can be joined along a $\phi$-string, with the phase of $\phi$
changing by $2\pi$ around the string.

If the field $a$ is
the phase of a complex scalar, $X=R e^{ia}$,  then the
translation symmetry (\ref{shift4}) is a $U(1)$ symmetry of phase
transformations, $X\rightarrow e^{i\alpha}X$.  When $X$ gets an
expectation value, this symmetry is broken and we expect global string
solutions with $a$ changing by $2\pi$ around a string.
An interesting question is what happens to these strings at the
$Z_{2N}$-breaking phase transition when $\phi$ gets an expectation
value (we assume that it does so after $X$).  With $\langle\phi
\rangle =const$, the field strength $F$ changes
around the string by the amount (\ref{DeltaF4}) with $\Delta a=2\pi$,
\beq
\Delta F=2\pi g\eta^2\epsilon^N,
\label{DF}
\eeq
suggesting that there should be a discontinuity
along a sheet attached to the string.  One way of resolving this
obstruction is to assume that each $X$-string gets covered by a
$\phi$-string with $2N$ walls attached to it, so that the cores of the
two strings coincide.\footnote{We thank J. Garriga for pointing this
out to us.}  As we go around such a combined string, the sign of
$\phi^N$ changes every time we cross a $\phi$-wall, and with it the
coupling between ${\tilde A}^\mu$ and $a$ in (\ref{L4}) also changes
sign.  As a result, the change $\Delta F$ will have different sign in
different sections between the $\phi$-walls, and the overall change
around the string will vanish.  Another possibility is that the field
$F$ changes by the amount (\ref{DF}) inside a wall which gets attached
to the $X$-string.  Within such $F$-walls, the field $\phi$ should
deviate from its vacuum value and the higher-derivative terms omitted
in the Lagrangian (\ref{L4}) should become important.  Which of the
two options is realized may depend on the specific dynamics of the
model.

Although the physics of topological defects in our model may be quite
interesting, these defects are probably irrelevant for cosmology.  The
reason is that $\phi$-walls would be disastrous if allowed to survive
until present.  The $Z_{2N}$ symmetry breaking should therefore occur
before the end of inflation, so that all $X$ and $\phi$ strings and walls
are inflated away and
we are left only with $a$-walls at the boundaries of nucleating
bubbles.

\section{Acknowledgments}

We are grateful to G.Gabadadze, J.Garriga and A. Grassi
for useful discussions.
The work of GD was supported in part by David and Lucille Packard Foundation
Fellowship for Science and Engineering, by Alfred P. Sloan foundation
fellowship and by the NSF grant PHY-0070787.  The work of A.V. was
supported in part by the NSF grant PHY-9601896.

\end{document}